\documentclass[12pt]{article}
\usepackage[utf8]{inputenc}
\usepackage{listings}
\usepackage[english]{babel}
\usepackage{xcolor}
\usepackage[hidelinks]{hyperref}
\usepackage[capitalise]{cleveref}
\usepackage{authblk}

\definecolor{codegreen}{rgb}{0,0.6,0}

\lstdefinestyle{mystyle}{
    commentstyle=\color{codegreen},
    keywordstyle=\color{blue},
    basicstyle=\ttfamily\footnotesize,
    breakatwhitespace=false,         
    breaklines=true,                 
    captionpos=b,                    
    keepspaces=true,                 
    numbersep=5pt,                  
    showspaces=false,                
    showstringspaces=false,
    showtabs=false,                  
    tabsize=2,
}

\newcommand{\cpp}{C++}
\newcommand{\vf}{VeriFast}
\newcommand{\clang}{C}
\newcommand{\java}{Java}

\title{Verification of C++ Programs with VeriFast}
\author{Niels Mommen}
\author{Bart Jacobs}
\affil{imec-DistriNet Research Group, KU Leuven, Belgium}
\affil{\{niels.mommen, bart.jacobs\}@kuleuven.be}
\date{December 2022}

\begin{document}
\emergencystretch 3em
\pagenumbering{Roman} 

\maketitle

\begin{abstract}
    \vf{} is a prototype tool based on separation logic for modular verification of \clang{} and \java{} programs. We are in the process of adding support for \cpp{}. In this report, we describe the features of \cpp{} for which we added support so far, as well as the proof obligations we generate for these features. At this point, \vf{} has basic support for most object-oriented programming features of \cpp{}: member functions, member function and operator overloading, implicit and explicit conversions, constructors and initializer lists, destructors, reference types, allocation and deallocation on the stack or on the heap (using \texttt{new} and \texttt{delete}), inheritance (including multiple inheritance but not virtual base classes), and virtual member functions and overriding. To support specification of inheritance hierarchies, we added support for instance predicates, which can be introduced in a base class and overridden in derived classes. The main missing feature at this point is support for \cpp{} templates, which we plan to work on next.
\end{abstract}

\cleardoublepage

\tableofcontents

\cleardoublepage

\pagenumbering{arabic}

\section{Introduction}
\vf{}\footnote{\url{https://github.com/verifast/verifast}} is a prototype tool that performs modular symbolic execution for modular verification of \clang{} and \java{} programs, based on separation logic \cite{jacobs2011verifast}. Currently, support for verification of programs written in \cpp{} is being added to \vf{}. We use LibTooling, a library to write tools based on Clang, to retrieve a typechecked abstract syntax tree for well-typed \cpp{} programs \footnote{\url{https://clang.llvm.org/docs/LibTooling.html}}. This report describes the \cpp{} features that are currently verifiable by \vf{} and the changes that were needed in order to verify these features. The added features include verification of member functions, constructors and destructors, and virtual member functions in the presence of multiple inheritance. To conclude, present limitations and future work is discussed. A snapshot of binaries and the source code of \vf{} with \cpp{} support at the time of writing is available as a Zenodo drop at \url{https://zenodo.org/record/7486648}.

\section{\cpp{} basic features}
A first extension added to \vf{} is support for lvalue references and the \texttt{new} and \texttt{delete} operator for primitive types. 

\paragraph{Lvalue references} \vf{} treats lvalue references similar to pointers. By default, an lvalue reference evaluates to the address the reference points to. When an lvalue to rvalue conversion is needed to retrieve to value of the object the reference points to, \vf{} implicitly dereferences the pointer. Lvalue reference types are currently not supported in ghost code. It is however possible to reason about an object that is referenced by an lvalue reference by taking the address of that reference, which is equivalent to a pointer to that object.

\paragraph{\texttt{new} and \texttt{delete}}
The \texttt{new} and \texttt{delete} operators in \cpp{} are the counterparts of \texttt{malloc} and \texttt{free} in \texttt{C}. However, it is not allowed to use them interchangeably: an object allocated on the heap through \texttt{new} cannot be destroyed by passing the address returned by \texttt{new} to \texttt{free}. Therefore, \vf{} respectively produces and consumes a \texttt{new\_block} when calling \texttt{new} and \texttt{delete}, instead of \texttt{malloc\_block} chunks.

\section{Classes and objects}
\label{sec:classes and objects}
Classes in \cpp{} are an extended form of structs in \clang{}. First, optional default expressions can be specified for data members. Such an expression is evaluated to initialize the data member when an instance of the class, an object, is created. 

During object creation of class type \texttt{S} at address \textit{addr}, \vf{} traverses each field of \texttt{S} in their order of declaration and produces a corresponding field chunk \texttt{S\_\textit{field\_name}(\textit{addr},\textit{value})}\footnote{A points-to chunk \texttt{\textit{a}->\textit{field\_name} |-> \textit{v}} can alternatively be used to refer to a field chunk at address \textit{a} with value \textit{v}.}, where \textit{field\_name} is the name of the class field and \textit{value} either is the evaluation result of \textit{field\_name}'s default member initializer or represents an unspecified value in case no initializer is present.

Next to data members, static and non-static member functions can be defined. A non-static member function always has an implicit \texttt{this} parameter, which is a pointer to an instance of the declaring class: it points to the target of the function call. 

When verifying a non-static member function, \vf{} produces a fresh symbol that represents its implicit \texttt{this} argument, which is assumed to not be zero, i.e. not a null pointer. Verification of a non-static member function continues as usual: first the precondition is produced, next the compound statement of the function body is verified, and finally the postcondition of the function is consumed and a check is performed to verify that no chunks are leaked. Calling a member function on a null pointer results in undefined behaviour. Therefore, a member function cannot be called when it not possible to prove that the target object of the member function call is not zero. 

\subsection{Construction of objects}
\label{subsec: construction of objects}
A constructor in \cpp{} is a special member function of a class that cannot be called directly; it is automatically called when an instance of that class is created. Multiple constructors can be defined within a class. An applicable constructor is selected using overload resolution when an object is created.

A constructor consists of an optional initializer list and a body consisting of a compound statement. During construction, all fields are initialized in their declaration order prior to executing the destructor's body. An initializer list optionally defines data member initializer expressions which take precedence over the default member initializers declared in its class. Therefore, during field initialization of data field \textit{m} of an object of class type \texttt{S}, \vf{} selects the appropriate expression from the initalizer list if it mentions \textit{m}. Otherwise, the default member initializer is used if available or
the field is initialized using its default constructor when the field type is a
class type, or it is initialized with an unspecified value if the field type is a
primitive type. Accordingly, verification of a constructor consists of the following steps:
\begin{enumerate}
    \item Initialize fields in order of declaration, where entries in the initializer list take precedence over default member initializers;
    \item Symbolically execute the constructor body.
\end{enumerate}

An initializer list of constructor \texttt{C} of class type \texttt{S} can alternatively consist of exactly one call to another constructor \texttt{C'}. This skips field initialization and delegates construction to the applicable constructor \texttt{C'} selected using overload resolution, in which case \vf{} verifies a call to \texttt{C'}. Afterwards, construction continues by verifying the body of \texttt{C}. 

An object can either be allocated on the stack or on the heap. A stack-allocated object of class type \texttt{S} is created when a variable of type \texttt{S} is declared by automatically verifying a call to an applicable constructor. Objects can be allocated on the heap by invoking a \texttt{new} expression. This expression invokes a call to an applicable constructor and returns a pointer to the object that has been allocated on the heap. Verification of a \texttt{new} expressions for an object of class type \texttt{S} first verifies a call to the constructor and additionally produces a \texttt{new\_block\_S(\textit{addr})} chunk analogously to a \texttt{malloc\_block\_S(\textit{addr})} chunk in \clang{}, where \textit{addr} is the address in the heap at which the object was created.

\subsection{Destruction of objects}
\label{subsec: destruction of objects}
Destructors are the counterpart of constructors: they are called automatically when the lifetime of an object ends. Contrary to constructors, destructors do not have an equivalent counterpart of an initializer list.

Verification of a destructor \texttt{D} for an object of class type \texttt{S} at address \textit{addr} happens in reverse order of verifying a constructor of \texttt{S}. First, the destructor body is verified. Next, all fields of the object are destroyed in reverse order of their declaration. For a field \texttt{m} of primitive type, its corresponding field chunk \texttt{S\_m(\textit{addr}, \_)} is consumed, where \texttt{\_} represents a dummy pattern which matches any value. Fields that are of class type \texttt{S'} are destroyed by verifying a call to the destructor of \texttt{S'}. Verification of a destructor performs following the steps:
\begin{enumerate}
    \item Symbolically execute the destructor body;
    \item Destruct fields in reverse order of declaration.
\end{enumerate}

Objects that are allocated on the stack are automatically destroyed at the end of their scope by verifying a call to their destructor. Explicitly calling a destructor on a stack-allocated object is not allowed by \vf{}. This would lead to undefined behaviour when an object gets destroyed automatically at the end of its scope when it was already explicitly destroyed.

For an object that was allocated on the heap at address \textit{addr}, the \texttt{delete} operator can be used to destruct its operand. Verification of a \texttt{delete \textit{addr}} expression, where \textit{addr} is a pointer to an object of class type \texttt{S}, takes place in two steps. First, a call to the destructor of \texttt{S} is verified. Afterwards, a \texttt{new\_block\_S(\textit{addr})} chunk is consumed. This guarantees that an object allocated on the heap is never destroyed twice, which would otherwise lead to undefined behaviour.

\section{Inheritance}
\label{sec: inheritance}
A class in \cpp{} can extend another class, inheriting all accessible members from the class it extends. When class \texttt{D} extends class \texttt{B}, \texttt{B} is called a (direct) base class of \texttt{D} and \texttt{D} is a derived class of \texttt{B}. 

\vf{} models a base object of class type \texttt{B} as a subobject in a derived object of class type \texttt{D}. For an object \textit{d} of class type \texttt{D} created at address \textit{d\_addr}, its base object of class type \texttt{B} can be accessed through a field pointer \texttt{field\_ptr(\textit{d\_addr}, D\_B\_offset)}, where \texttt{D\_B\_offset} represents an offset, $\geq 0$, from \textit{d} to its base object of class type \texttt{B}.

\subsection{Upcasts}
\label{subsec: upcasts}
Upcasting a derived object to its base object is supported by \vf{} both implicitly and explicitly. Such a cast is first required when accessing a base field member. Second, a cast is needed when passing a derived object as an argument to a function or method parameter that expects a base object, base object pointer or reference. Lastly, casting a derived object to its base object is required when calling a base member function on a derived object. 

The evaluation of an upcast from a derived object of class type \texttt{D} at address \textit{d\_addr} to a base object of class type \texttt{B} is executed in two steps. First a check is performed to make sure that \texttt{D} derives from \texttt{B}. Next, a field pointer \texttt{field\_ptr(\textit{d\_addr}, D\_B\_offset)} is computed to retrieve the address of the base object.

Note that implicit upcasts are limited in ghost code. The type checker of \vf{} automatically inserts them when it is able to deduce the need for such a cast. This is currently not supported when a points-to notation is used to reason about base fields in a derived object. E.g., \texttt{B\_m(\textit{d\_addr, ?m})} implicitly involves an upcast to access base field \textit{m} from derived object \textit{d}, while \texttt{((B *) d\_addr)->m |-> ?m} requires an explicit upcast to access the same field.

\subsection{Multiple inheritance}
\label{subsec: multiple inheritance}
\cpp{} offers the feature to derive from multiple base classes. Treating base classes as subobjects through field pointers in \vf{} includes support for multiple inheritance. Fields of any base object can be accessed by first upcasting the derived object to the base, prior to accessing the field in that base. Similarly, visible base member functions from any base can be used through a derived object by first implicitly or explicitly performing an upcast to the base target.

In the presence of multiple inheritance, an implicit upcast from a derived object to a base object might be ambiguous when multiple base classes derive from the same class. Assume a diamond scenario where class \texttt{D} derives from both class \texttt{B} and class \texttt{C}, and classes \texttt{B} and \texttt{C} in turn both derive from class \texttt{A}. In this scenario, an object of class type \texttt{D} at address \textit{d\_addr} has two subobjects: one object of class type \texttt{B} at address \texttt{field\_ptr(\textit{d\_addr}, D\_B\_offset)} and another object of class type \texttt{C} at address \texttt{field\_ptr(\textit{d\_addr}, D\_C\_offset)}. These subobjects both have one subobject of class type \texttt{A} themselves at respectively address \texttt{field\_ptr(field\_ptr(\textit{d\_addr}, D\_B\_offset), B\_A\_offset)} and address \texttt{field\_ptr(field\_ptr(\textit{d\_addr}, D\_C\_offset), C\_A\_offset)}. Accessing a field \textit{m} of class \texttt{A} from \textit{d\_addr} is ambiguous: it not clear whether this refers to field \textit{m} reachable through the subobject of class type \texttt{B} or the subobject of class type \texttt{C}. Hence, an explicit cast is required to render the upcast unambiguous. E.g., \texttt{A\_m((B *) \textit{d\_addr}, ?m\_val)} can be used to reason about field \textit{m} in the subobject at address \texttt{field\_ptr(field\_ptr(\textit{d\_addr}, D\_B\_offset), B\_A\_offset)}.

\subsection{Construction and destruction}
Verification of constructors and destructors in the presence of inheritance requires additional steps in the verification process. Verifying a constructor first starts with constructing all base classes in declared derivation order if the constructor being verified is not a delegating constructor. Next, verification proceeds as described before in \cref{subsec: construction of objects}. 

\vf{} verifies the construction of bases by verifying a call to a base constructor of each direct base class. An appropriate base constructor is selected by overload resolution if the initializer list of the constructor mentions a base class initialization, otherwise the default base constructor is used. The value of the implicit \textit{this} argument passed to the base constructor is calculated by performing an upcast from the current constructor's implicit \textit{this} parameter to the base object that will be constructed by verifying the selected constructor call. E.g., when verifying a base constructor call of class \texttt{B} in the constructor of derived class \texttt{D} for an object at address \textit{d\_addr}, a value \texttt{field\_ptr(\textit{d\_addr}, D\_B\_offset)} is passed as the implicit \textit{this} argument to the constructor call of class \texttt{B}.

Verification of a destructor accordingly accounts for inheritance. After following the verification steps for destruction listed in \cref{subsec: destruction of objects}, \vf{} first verifies a call to each destructor of all direct bases in reverse order of derivation if the class object to be destroyed has any base objects. Analogously to constructors, an upcast to the base object is performed when calculating the value passed as the implicit \textit{this} argument to the base destructor verification call.

Care has to be taken during construction and destruction of objects in the presence of inheritance. Member functions can be called directly and indirectly during construction or destruction, but this results in undefined behaviour if not all bases have been fully constructed. Therefore, an \texttt{S\_bases\_constructed(S *\textit{s\_addr})} chunk is introduced to determine whether all bases for an object of class type \texttt{S} at address \textit{s\_addr} have been constructed. This chunk is required when verifying a member function call where the target object derives from at least one class, disallowing calling any member function when this chunk is not available.

Verification of a constructor can now be summarized by the following steps:
\begin{enumerate}
    {\color{blue}
    \item Verify a constructor call for each base class in order of derivation;
    \item Produce an \texttt{S\_bases\_constructed(\textit{s\_addr})} chunk if \texttt{S} derives from at least one class, where \texttt{S} is the class type of the object that is currently being constructed and \textit{s\_addr} is the address of the object;
    }
    \item Initialize fields in order of declaration, where entries in the initializer list take precedence over default member initializers;
    \item Symbolically execute the constructor body.
\end{enumerate}

Verification of a destructor now additionally destructs base classes and optionally consumes a \texttt{bases\_constructed} chunk:
\begin{enumerate}
    \item Symbolically execute the destructor body;
    \item Destruct fields in reverse order of declaration;
    {\color{blue} 
    \item Consume an \texttt{S\_bases\_constructed(\textit{s\_addr})} chunk if \texttt{S} derives from at least one class, where \texttt{S} is the class type of the object that is currently being destructed and \textit{s\_addr} is the address of the object;
    \item Verify a destructor call for each base class in reverse order of derivation.
    }
\end{enumerate}

\section{Virtual methods}
A Member function can be declared \textit{virtual} with the \texttt{virtual} keyword. This allows to override the member function in derived classes. Dynamic dispatch is used to select the appropriate member function implementation when an unqualified member function call is evaluated: the dynamic type of the target object is inspected at run time in order to select a member function implementation. That is, if a virtual member function is called on a base object, dynamic dispatch selects the \textit{final overrider}. A virtual member function in a base class is a final overrider if no derived class declares a member function that overrides it. Qualified member function calls are interpreted as statically bound calls, not resulting in dynamic dispatch at run time. A class (object) that has at least one virtual member function, will be referred to as a \textit{polymorphic} class (object).

\subsection{Object types at run time}
\label{subsec:runtime types}
In order to reason about virtual member functions and polymorphic classes, \vf{} first introduces a \texttt{typeid} construct that can be used in ghost code. This construct acts similar to the \texttt{typeid} operator in \cpp{}: it takes one argument, a type expression, and returns a reference to an \texttt{std::type\_info} object which uniquely represents that type. The difference is that \texttt{typeid} in \vf{} only accepts type expressions as its argument, whereas the \texttt{typeid} operator in \cpp{} accepts any expression. A \texttt{typeid(S)} evaluates to value \texttt{S\_type\_info}, where \texttt{S} is a type expression referring to class type \texttt{S}.

In addition, for each polymorphic class \texttt{S}, \vf{} introduces a predicate \texttt{S\_vtype(S *\textit{s\_addr}; std::type\_info *s\_info)}. This predicate allows to reason about the run-time type or its \textit{most derived} type. In case class \texttt{S} has at least one polymorphic base, this predicate is defined as 
\begin{lstlisting}[language=C++]
/*@ predicate S_vtype(S *s_addr, std::type_info *s_info) =
    B0_vtype(s_addr, s_info) &*&
    B1_vtype(s_addr, s_info) &*&
    ... &*&
    Bn_vtype(s_addr, s_info);
@*/
\end{lstlisting} where $B_0,\ldots,B_n$ are polymorphic direct base classes of \texttt{S} with $n>0$. Otherwise, the \texttt{vtype} predicate is opaque and cannot be opened or closed.

This allows to open a \texttt{vtype} chunk of a polymorphic class in order to retrieve all \texttt{vtype} chunks of its polymorphic direct bases. Analogously, all \texttt{vtype} chunks of polymorphic direct base objects are required to close a \texttt{vtype} chunk of a polymorphic derived object.

When verifying a virtual member function call on a target object at address \textit{s\_addr} with static class type \texttt{S}, \vf{} first checks that (some fraction of) an \texttt{S\_vtype(\textit{s\_addr}, \_)} chunk is available. Otherwise, the member function call is not allowed. In order to call a base member function on a derived target object, the \texttt{vtype} chunk of the derived object can be opened to obtain a \texttt{vtype} chunk for the base object.

\subsection{Construction and destruction}
To be able to call virtual member functions, \texttt{vtype} chunks for polymorphic objects have to be produced at some point during construction. These chunks then have to be consumed during destruction of an object.

Like regular member functions, virtual member functions can be called during construction and destruction of an object. However, in presence of multiple inheritance, care has to be taken. First, during construction or destruction of an object, the virtual function called is the final overrider in the constructor or destructor class. Hence, potential overriders in derived classes are not taken into account during the construction or destruction of a base class. 

A second point of attention is the presence of multiple inheritance. During construction and destruction, virtual base member function calls on the object under construction or destruction are only allowed for bases that belong to the inheritance sub hierarchy of that object. Otherwise, if a virtual member function is called on a subobject belonging to another branch of the inheritance hierarchy, the virtual member call would result in undefined behaviour.
\begin{figure}
\begin{lstlisting}[language=C++, caption={Virtual member call during construction, resulting in undefined behaviour.}, label={lst: vcall ub}]
struct A {
  virtual void foo() {}
};

struct B {
  B(A *a) {
    a->foo(); // undefined behaviour
  }

  virtual void bar() {}
};

struct C : public A, public B {
  C() : A(), B(this) {
    foo();
    bar();
  }
};
\end{lstlisting}
\end{figure}

\Cref{lst: vcall ub} illustrates an example where virtual base member functions are called in the constructor of \texttt{C}. These calls are allowed because at the time of the call, both base \texttt{A} and \texttt{B} are fully constructed and they belong to the inheritance hierarchy of \texttt{C}. However, the constructor of \texttt{B} that is called through the constructor of class \texttt{C} would lead to undefined behaviour because it calls member function \texttt{foo} of base class \texttt{A}: the subobject of class type \texttt{A} is fully constructed, but it does not belong to the inheritance hierarchy of the subobject of class type \texttt{B}.

In order to address these instances where calling virtual member functions would lead to undefined behaviour, \vf{} produces the \texttt{D\_vtype(\textit{d\_addr}, D\_type\_info)} chunk in the constructor of polymorphic class \texttt{D} after all its bases have been constructed, where \textit{d\_addr} is the address of the object that is constructed. Right after a constructor of a polymorphic base class \texttt{B} has finished, the derived constructor of class \texttt{D} consumes the \texttt{B\_vtype(field\_ptr(\textit{d\_addr}, D\_B\_offset), B\_type\_info)} that was produced by the base constructor call. This disallows other base objects that still have to be constructed in other branches of the inheritance hierarchy from calling virtual methods in branches that would lead to undefined behaviour. Verification of a constructor goes as follows:
\begin{enumerate}
    \item Verify a constructor call for each base class in order of derivation;
    {\color{blue}
    \begin{enumerate}
        \item If the base class is polymorphic, consume a \texttt{B\_vtype(field\_ptr( \linebreak\textit{d\_addr} D\_B\_offset), B\_vtype\_info)} chunk, where \textit{d\_addr} is the address of the derived object of class type \texttt{D} and \texttt{B} represents the class type of the base object;
    \end{enumerate}
    \item Produce a \texttt{D\_vtype(\textit{d\_addr}, D\_type\_info)} chunk if the object under construction is polymorphic, where \texttt{D} is the class type of the object under construction;
    }
    \item Produce a \texttt{D\_bases\_constructed(\textit{d\_addr})} chunk if \texttt{D} derives from at least one class, where \texttt{D} is the class type of the object that is currently being constructed and \textit{d\_addr} is the address of the object;
    \item Initialize fields in order of declaration, where entries in the initializer list take precedence over default member initializers;
    \item Symbolically execute the constructor body.
\end{enumerate}

Verification of a destructor again changes in a similar fashion. After destructing all member fields of class \texttt{D}, \vf{} consumes a \texttt{vtype(\textit{d\_addr}, D\_type\_info)} chunk if it is polymorphic, where \texttt{d\_addr} is the address of the object under destruction. This disables polymorphic base objects from calling virtual member functions of its derived object and calling member functions of bases that are are part of another branch in the inheritance hierarchy. Prior to verifying a polymorphic base destructor call, \vf{} produces a \texttt{B\_vtype(\textit{b\_addr}, B\_type\_info)} chunk, where \texttt{B} is the class type of the base object at address \textit{b\_addr}. This allows the polymorphic base object to call virtual member functions in its inheritance sub hierarchy. Hence, verification of a destructor can be performed through the following steps:
\begin{enumerate}
    \item Symbolically execute the destructor body;
    \item Destruct fields in reverse order of declaration;
    {\color{blue}
    \item If the object of class type \texttt{D} at address \textit{d\_addr} under destruction is polymorphic, consume a \texttt{D\_vtype(\textit{d\_addr}, D\_type\_info)} chunk;
    }
    \item Consume a \texttt{D\_bases\_constructed(\textit{d\_addr})} chunk if \texttt{D} derives from at least one class, where \texttt{D} is the class type of the object that is currently being constructed and \textit{d\_addr} is the address of the object;
    \item Verify a destructor call for each base class in reverse order of derivation;
    {\color{blue}
    \begin{enumerate}
        \item Prior to verifying a base destructor call of a base object of class type \texttt{B} that is polymorphic, produce a \texttt{vtype(field\_ptr(\textit{d\_addr}, D\_B\_offset), B\_type\_info)}, where \texttt{D} is the class type of the derived object at address \textit{d\_addr}.
    \end{enumerate}
    }
\end{enumerate}

\subsection{Behavioural subtyping}
\label{subsec: behavioral subtyping}
When verifying a virtual member function call, \vf{} uses the same contract that would be used during a statically bound call. However, the run-time type of the target object might be a derived type of the statically known target type. The static type acts as an upper bound for the objects's dynamic type \cite{leavens2000concepts}. Therefore, \vf{} performs a behavioural subtyping check for each virtual member function that is overridden to make sure that the overriding member function can statically call the member function that is overridden \cite{parkinson2005local}. First, the precondition of the overridden member function is produced. Next, \vf{} consumes the precondition of the overriding member function and produces the postcondition of the overriding member function. Lastly, \vf{} consumes the postcondition of the overridden member function.

This behavioural subtyping requirement makes verification of virtual member functions harder and sometimes impossible. It often also requires changes to the specification of the member function that is overridden and reverification of the overridden member function. 

\begin{figure}
\begin{lstlisting}[language=C++, caption={Virtual member function override, violating behavioural subtyping.}, label={lst: behavioral subt violation}]
struct A {
  virtual int foo()
  //@ requires true;
  //@ ensures result >= 0;
  {
    return 0;
  }
};

struct B : public A {
  int m = 10;

  int foo() override
  /*@ requires 
      B_bases_constructed(this) &*& 
      this->m |-> ?m_val &*&
      m_val >= 0; 
  @*/ 
  /*@ ensures 
      B_bases_constructed(this) &*& 
      this->m |-> m_val &*& 
      result == m_val; 
  @*/
  {
    return m;
  }
};
\end{lstlisting}
\end{figure}

Take as an example the code snippet in \cref{lst: behavioral subt violation}. Class \texttt{B} overrides member function \texttt{foo} from its base class \texttt{A}. The specification in class \texttt{B} clearly violates the behavioural subtyping check: the precondition of the overriden member function does not imply the precondition of the overriding member function. In order to be able to verify this program, the precondition of \texttt{foo} in \texttt{A} has to be changed to require the chunks mentioned in the precondition of \texttt{foo} in \texttt{B}. However, these chunks would only be available when the run time class type of the target object would be \texttt{B}. Changing the specification requires reverification of the overridden method, making modular verification harder. In order to tackle this problem, \vf{} supports the use of instance predicates.

\section{Instance predicates}
\vf{} already supports dynamically bound instance predicates for \java{} programs \cite{smans2013verifast}. An instance predicate can be defined in a class body and does not have a single definition: each subclass defines its own definition of the predicate. We also added this feature for classes in \cpp{}.

\begin{figure}
\begin{lstlisting}[language=C++, caption={Shape and Square example with instance predicates.}, label={lst: inst pred ex}]
struct Shape {
  //@ predicate valid() = true;

  Shape()
  //@ requires true;
  //@ ensures valid() &*& Shape_vtype(this, thisType);
  {
    //@ close valid();
  }

  virtual int calcArea() const = 0;
  //@ requires valid();
  //@ ensures valid() &*& result >= 0;
};

class Square : public Shape {
  int m_width;

public:
  /*@ predicate valid() =
      this->valid(&typeid(Shape))() &*&
      this->m_width |-> ?w &*&
      w >= 0 &*&
      Square_bases_constructed(this);
  @*/

  Square(int width) : m_width(width)
  //@ requires width >= 0;
  //@ ensures valid() &*& Square_vtype(this, thisType);
  {
    //@ close valid();
  }

  int calcArea() const override
  //@ requires valid();
  //@ ensures valid() &*& result >= 0;
  {
    //@ open valid();
    return m_width * m_width;
    //@ mul_mono_l(0, this->m_width, this->m_width);
    //@ close valid();
  }
};
\end{lstlisting}
\end{figure}

An instance predicate has, similar to \cpp{} member functions, an implicit \textit{this} parameter that is bound to the target object. Each derived class can override an instance predicate that is declared in one of its base classes. Hence, an instance predicate chunk contains an index argument to distinguish between the different versions of the predicate. This index is a pointer to an \texttt{std::type\_info} object in \vf{} when verifying \cpp{} programs. For example, an index with value \texttt{S\_type\_info} determines the instance predicate defined in class \texttt{S}. Every type has a unique \texttt{std::type\_info} object, which allows to differentiate overridden instance predicate chunks by their index.

The example in \cref{lst: inst pred ex} defines an abstract \texttt{Shape} class that declares an instance predicate \texttt{valid}. This class is inherited by the \texttt{Square} class, which also overrides the definition of \texttt{valid}. An instance predicate chunk of \texttt{valid} in \texttt{Shape} has signature \texttt{Shape\#valid(\textit{addr}, Shape\_type\_info)}, where \textit{addr} is a pointer to the target object and \texttt{Shape\_type\_info} is the symbolic value that represents a pointer to the \texttt{std::type\_info} object of class \texttt{Shape}. Remember that this value can be obtained by evaluating \texttt{\&typeid(Shape)}.

Both the constructor of \texttt{Shape} and \texttt{Square} in \cref{lst: inst pred ex} mention a \textit{thisType} variable: a ghost variable that exists to make verification feasible. This variable is implicitly available in each non-static member function of a class and represents the value that would be obtained from evaluating \texttt{\&typeid(S)}, where \texttt{S} is the class type of the target object that is bound the member function call. The \textit{thisType} ghost variable is implicitly used by \vf{} as the index of an instance predicate where the target object is implicit \textit{this}. Hence, the chunk produced by \texttt{valid()} in the postcondition of constructor \texttt{Shape()} is \texttt{Shape\#valid(\textit{addr}, \textit{thisType})}, where \textit{addr} is the address of the \texttt{Shape} object that would be constructed by the constructor.

The value of \textit{thisType} depends on whether a member function call is dynamically or statically bound. This allows to use a different interpretation of the contract for statically bound member function calls and dynamically bound member function calls, only requiring one specification from the programmer. I.e., the value of \textit{thisType} depends on the binding of a member function call.

The value of \textit{thisType} is assumed to be equal to \texttt{S\_type\_info} during verification of a non-static member function, where \texttt{S} is the static type of the target object. On the other hand, during verification of a dynamically bound member function call, \textit{thisType} is assumed to be a pointer to the \texttt{std::type\_info} object for the most derived class type of the target. As explained in \cref{subsec:runtime types}, a check is performed for the existence of (some fraction of) an \texttt{S\_vtype(\textit{addr}, \textit{?info})} before verifying a virtual member function call, where \texttt{S} is the static class type of the target expression and \textit{addr} its address. \vf{} now assumes that \textit{thisType} has value \textit{info} during verification of the virtual member function call. The interpretation of a specification must be the same during verification at the call site and during verification of the callee. Therefore, \vf{} checks that a derived class overrides all virtual member functions in all of its polymorphic direct bases to meet this requirement \cite{barnett2004verification}.

The target object of a predicate instance assertion can also be mentioned explicitly, as \cref{lst: inst pred ex} shows in the instance predicate definition of \texttt{valid} in class \texttt{Square}. In this instance, \texttt{this->valid(\&typeid(Shape))()} refers to the instance predicate definition of \texttt{Shape} and is represented by a \texttt{Shape\#valid(this, Shape\_type\_info)} chunk.

Evaluation of an instance predicate assertion with an explicit target object that does not mention an explicit index depends on the nature of its target. For a class \texttt{S} that is not polymorphic and defines an instance predicate \texttt{s\_pred($a_0$,$\ldots$,$a_n$)}, and an object of class type \texttt{S} at address \textit{s\_addr}, \texttt{s\_addr->s\_pred($a_0$,$\ldots$,$a_n$)} evaluates to \texttt{S\#s\_pred(s\_addr, S\_type\_info, $a_0$,$\ldots$,$a_n$)}. However, when \texttt{S} is polymorphic, \vf{} treats \texttt{s\_addr->s\_pred($a_0$,$\ldots$,$a_n$)} as syntactic sugar for \texttt{S\_vtype(s\_addr, ?s\_type) \&*\& s\_addr->s\_pred(s\_type)($a_0$,$\ldots$,$a_n$)}.

\begin{figure}
\begin{lstlisting}[language=C++, caption={Verifiable version of \cref{lst: behavioral subt violation} with instance predicates, while preserving behavioural subtyping.}, label={lst: inst pred beh subt}]
struct A {
  //@ predicate valid() = true;

  virtual int foo()
  //@ requires valid();
  //@ ensures valid() &*& result >= 0;
  {
    return 0;
  }
};

struct B : public A {
  int m = 10;
  
  /*@ predicate valid() =
      this->m |-> ?m_val &*&
      m_val >= 0 &*&
      B_bases_constructed(this);
  @*/

  int foo() override
  //@ requires valid();
  //@ ensures valid() &*& result >= 0;
  {
    //@ open valid();
    return m;
    //@ close valid();
  }
};
\end{lstlisting}
\end{figure}

Using instance predicates, we can now annotate and verify the example from \cref{lst: behavioral subt violation} while preserving behavioural subtyping as illustrated in \cref{lst: inst pred beh subt}.

\section{Limitations}
\label{sec:limitations}
It is currently only possible to pass objects by reference or by pointers. No support has been added yet to pass objects by value. Therefore, all function parameters and function return types are either primitive types, reference to objects or pointers to objects.

Explicit destructor calls are currently not allowed by \vf{}. This avoids explicitly destroying a stack-allocated object, when it might later be destroyed automatically when the object goes out of scope, potentially leading to undefined behaviour. However, explicit constructor calls are required once placement \texttt{new} is supported in order to construct an object in preallcoated memory.

Verification for virtual destructors is infeasible when a polymorphic class inherits from multiple polymorphic base classes that have a virtual destructor. The overriding virtual destructor in the derived class can only meet the requirement of behavioural subtyping for one of its base classes, because it is currently not supported to declare an instance predicate in a derived class that overrides multiple instance predicates defined in its base classes. 

Furthermore, heap-allocated objects that inherit from at least one base object cannot be destroyed with the \texttt{new} operator through a pointer to one of its base objects, even if it has a virtual destructor. Verification of the \texttt{delete} operator on a pointer to a base object of a derived object requires a \texttt{new\_block} chunk that was allocated for the base object.

\section{Future work}
Support for verification of \cpp{} programs with \vf{} is still in development. We are first planning to address the limitations discussed in \cref{sec:limitations}. In addition, we plan to add support for virtual inheritance and rvalue references. Next, our goal is support for verification of programs with \cpp{} templates, which is one of the main missing features in \vf{}. \textit{Constraints} and \textit{concepts}, language features added in \cpp{}20, are interesting aspects when looking into verification of programs with templates. Once these features have been added to \vf{}, we plan to verify some critical industrial \cpp{} programs. This would give insight in \cpp{} features that are commonly used and might involve interesting and new verification challenges.

\cleardoublepage
\bibliographystyle{plain}
\bibliography{refs}

\begin{thebibliography}{1}

\bibitem{barnett2004verification}
Michael Barnett, Robert DeLine, Manuel F{\"a}hndrich, K~Rustan~M Leino, and
  Wolfram Schulte.
\newblock Verification of object-oriented programs with invariants.
\newblock {\em J. Object Technol.}, 3(6):27--56, 2004.

\bibitem{jacobs2011verifast}
Bart Jacobs, Jan Smans, Pieter Philippaerts, Fr{\'e}d{\'e}ric Vogels, Willem
  Penninckx, and Frank Piessens.
\newblock {VeriFast}: A powerful, sound, predictable, fast verifier for {C} and
  {Java}.
\newblock In {\em NASA formal methods symposium}, pages 41--55. Springer, 2011.

\bibitem{leavens2000concepts}
Gary~T Leavens, Krishna~Kishore Dhara, and Krishna~Kishore Dhara.
\newblock Concepts of behavioral subtyping and a sketch of their extension to
  component-based systems.
\newblock 2000.

\bibitem{parkinson2005local}
Matthew~J Parkinson.
\newblock Local reasoning for {Java}.
\newblock Technical report, University of Cambridge, Computer Laboratory, 2005.

\bibitem{smans2013verifast}
Jan Smans, Bart Jacobs, and Frank Piessens.
\newblock {VeriFast} for {Java}: A tutorial.
\newblock {\em Aliasing in Object-Oriented Programming. Types, Analysis and
  Verification}, pages 407--442, 2013.

\end{thebibliography}

\end{document}